\documentclass[a4paper,11pt]{article}
\usepackage[german,american]{babel}
\usepackage[a4paper,top=3.5cm,bottom=3.5cm,left=3.0cm,right=3.0cm,marginparwidth=1.75cm]{geometry}
\usepackage{amsmath}
\usepackage{amsfonts}
\usepackage{amssymb}
\usepackage{graphicx}
\usepackage[font=small,labelfont=bf]{caption}
\usepackage{subcaption}
\usepackage{float}
\usepackage[OMLmathsfit]{isomath}
\usepackage{amssymb} 
\usepackage{amsxtra} 
\usepackage{bm} 
\usepackage{mathtools}
\usepackage[usenames]{color}
\usepackage{verbatim}
\usepackage[colorlinks=true, allcolors=blue]{hyperref}
\usepackage{soul}
\usepackage{ulem}
\pdfoutput=1

\DeclareCaptionLabelFormat{mylabel}{#1 #2\hspace{1.5ex}}
\usepackage[numbers,sort&compress]{natbib}
\usepackage{xcolor}
\usepackage{color}
\definecolor{darkgreen}{rgb}{0.0,0.5,0.0}
\definecolor{BurntOrange}{rgb}{0.8,0.3,0.0}
\definecolor{mygray}{gray}{0.5}
\definecolor{darkred}{rgb}{0.8,0,0}
\definecolor{darkblue}{rgb}{0,0,0.8}
\definecolor{gray}{rgb}{0.8,0.8,0.8}
\definecolor{mauve}{rgb}{0.58,0,0.82}

\usepackage{booktabs}

\providecommand{\keywords}[1]
{
  {
  \small	
  \textbf{\textit{Keywords---}} #1
}}

\definecolor{Iman}{RGB}{255,0,0}   
\definecolor{author2color}{RGB}{0,255,0}   
\definecolor{author3color}{RGB}{0,0,255}   

\title{Teaching Artificial Intelligence to Perform Rapid, Resolution-Invariant Grain Growth Modeling via Fourier Neural Operator}
\author{Iman Peivaste$^{1,2,*}$, Ahmed Makradi$^{1}$, 
Salim Belouettar$^{1}$ \\
\\
        \footnotesize $^1$Luxembourg Institute of Science and Technology (LIST), 5, Avenue des Hauts-Fourneaux, \\
        \footnotesize Esch-sur-Alzette, 4362, Luxembourg\\ 
       \footnotesize $^2$University of Luxembourg, 1511 Luxembourg, Luxembourg
 \\ 
        \footnotesize $^*$Corresponding Authors: iman.peivaste@list.lu
} 

\date{November 2024}

\begin{document}
\parskip4pt
\parindent0pt

\maketitle
\textbf{Accepted for Publication:} \\
This manuscript has been accepted for publication in \textit{Computer Methods in Applied Mechanics and Engineering}. The final version will be available via DOI upon publication.
\begin{abstract} 
Microstructural evolution, particularly grain growth, plays a critical role in shaping the physical, optical, and electronic properties of materials. Traditional phase-field modeling accurately simulates these phenomena but is computationally intensive, especially for large systems and fine spatial resolutions. While machine learning approaches have been employed to accelerate simulations, they often struggle with resolution dependence and generalization across different grain scales.
This study introduces a novel approach utilizing Fourier Neural Operator (FNO) to achieve resolution-invariant modeling of microstructure evolution in multi-grain systems. FNO operates in the Fourier space and can inherently handle varying resolutions by learning mappings between function spaces. By integrating FNO with the phase-field method, we developed a surrogate model that significantly reduces computational costs while maintaining high accuracy across different spatial scales.
We generated a comprehensive dataset from phase-field simulations using the Fan–Chen model, capturing grain evolution over time. Data preparation involved creating input-output pairs with a time shift, allowing the model to predict future microstructures based on current and past states. The FNO-based neural network was trained using sequences of microstructures and demonstrated remarkable accuracy in predicting long-term evolution, even for unseen configurations and higher-resolution grids not encountered during training.
\end{abstract}
\keywords{Machine learning, Deep learning, Fourier Neural Operator (FNO), Phase-Field, Grain Growth, Allen-Cahn, Fan-Chen}
\noindent
\section{Introduction}\label{sec1}

In materials science, the formation and evolution of complex microstructures play a critical role in determining the physical properties of materials. Grains with various orientations or domains with distinct structural variations can form complex patterns that evolve over time under the influence of driving forces such as temperature or stress \cite{biner2017overview, miyoshi2021large,flint2019phase}. Controlling microstructure evolution, such as grain growth, is essential for tailoring material properties to meet specific engineering needs \cite{yang2021phase}. For instance, adjusting grain size and its evolution can enhance mechanical properties like strength and toughness in metallic alloys \cite{hashmi2014comprehensive, ZOLLNER2016}, or, more recently, improve the performance of materials in emerging energy applications. In response to the growing need for clean energy, photoelectrochemical cells (PECs) used for water splitting (PEC-WS) have gained significant attention for solar-to-hydrogen conversion \cite{jiang2017photoelectrochemical}. The microstructure of materials in PECs affects charge transport, surface reaction kinetics, and light absorption properties, all of which are critical for the performance of these cells. Understanding and optimizing microstructural evolution enable the design of photoelectrodes with tailored properties, leading to enhanced cell efficiency and durability in new systems for clean energy \cite{tan2019fullerene, yang2021crucial}.

Due to its significant impact on material performance, microstructure evolution has been extensively studied across multiple scales, from atomistic to continuum \cite{, Haslam2001MechanismsOG, hashmi2014comprehensive,ZOLLNER2016,najafkhani2021recent, Gao2020ModesOG, louat1974theory, Trautt2012CoupledMO, Trautt2012GrainBM, kang2004sintering, wei2017three,okita2018grain, lopez2017two, ravash2017three, korbuly2017orientation}. Among various simulation techniques, the phase-field method stands out as a versatile and robust approach for accurately representing the physics of microstructure evolution over realistic timeframes \cite{flint2019phase, yongbo_liu__2024, yijing_shang__2023}. In the phase-field approach, grain boundaries move in response to applied forces, aiming to minimize the system's total free energy.
In this approach, phases, microstructures, and chemical compositions are represented by some field variables that transition smoothly across interfaces. The material's energy landscape, which determines the local driving forces, is defined by these phase fields or, in the case of non-conservative phase fields, by order parameters. Hence, phase-field models are generally categorized as either Cahn–Hilliard (CH) models \cite{cahn1958free, cahn1961spinodal}  for conservative phase fields (such as chemical concentration) or Allen–Cahn (AC) models \cite{allen1972ground, allen1973correction}  for non-conservative phase fields (such as structural order parameters).
However, despite its predictive power, phase-field modeling is computationally expensive \cite{flint2019phase, montes2021accelerating, ZHANG2020113362,https://doi.org/10.48550/arxiv.2205.02121, peivaste2022machine, oommen2022learning, bonneville2024accelerating}, especially when simulating large representative volume elements (RVEs) that encompass numerous grains. Solving the associated initial boundary value problems (IBVPs) for such systems often becomes impractical, particularly when complex models and large-scale simulations are involved. The high computational cost is further exacerbated in cases requiring fine spatial and temporal resolution, as well as the coupling of multiple physical fields (e.g., stress, and temperature). To alleviate these computational challenges, recent advances have focused on leveraging high-performance computing \cite{Sakane2015GPUaccelerated3P, miyoshi2017ultra}, optimizing numerical algorithms \cite{CHENG201545}, and incorporating machine learning (ML) techniques to accelerate simulations. In particular, surrogate models based on neural networks have shown promise in reducing computation time by orders of magnitude \cite{mianroodi2021teaching,nyshadham2019machine,wang2018multiscale}. Deep learning approaches such as deep neural networks (DNNs) \cite{zhang2020machine}, convolutional neural networks (CNNs) \cite{peivaste2022machine, CHOI2024103938}, recurrent neural networks (RNNs) \cite{montes2021accelerating, yang2021self}, and transfer learning \cite{ALHADALAHBABI2024117167} have been employed to approximate the solutions of phase-field models. However, these methods often struggle with generalization across different resolutions and grain scales, limiting their applicability to real-world materials systems where resolution invariance is critical.

Recently, neural operators have gained prominence as a new class of surrogate models for partial differential equations (PDEs). Unlike conventional neural networks that are dependent on specific resolutions, neural operators learn mappings between function spaces. This capability enables them to perform tensor-to-tensor regression that is invariant to resolution. Notable architectures in this category include the Fourier Neural Operator (FNO) \cite{li2020fourier}, Deep Operator Networks (DeepONets) \cite{lu2019deeponet},

An example of this approach is the work by Oommen et al. \cite{oommen2022learning}, who developed a framework based on DeepONet. This framework combines a convolutional autoencoder with a deep neural operator to predict microstructure evolution by learning the dynamic behavior of a two-phase mixture. Recently, Bonneville et al. \cite{bonneville2024accelerating} adapted a U-shaped adaptive Fourier Neural Operator (U-AFNO) to map conserved field variables from one time step to the next.

Building on these advances, this study introduces a novel approach using an adjusted FNO-based model to achieve resolution-invariant analysis in multi-grain systems where field variables are non-conserved. By integrating FNO with the non-conservative phase fields, we develop a resolution-invariant model capable of simulating grain evolution with significantly reduced computational costs, while maintaining accuracy across different spatial scales.
This study is an extension of our previous research \cite{peivaste2022machine}, in which we trained a U-Net based on CNNs to simulate grain growth in a dual-phase system. However, that system was resolution-dependent; for each resolution, we needed to train and tune a new network. Moreover, as most phase-field problems have periodic boundary conditions, standard CNNs struggle to capture this essential feature, leading to high errors at the boundaries. In contrast, FNO are inherently capable of handling periodic boundary conditions due to their operation in Fourier space, which naturally incorporates periodicity. This characteristic of FNO helps to significantly reduce boundary errors and enhances the model's ability to generalize across different resolutions. In this study, we demonstrate that the FNO can effectively address resolution-related challenges, offering a scalable and efficient approach for simulating grain evolution in AC-type phase-field models. The proposed FNO-based model not only significantly reduces computational costs but also enhances generalization across different grain scales, making it applicable to a wider range of material systems. This advancement represents a significant step forward in accelerating phase-field simulations and opens new avenues for integrating artificial intelligence with computational materials science to design and optimize materials with tailored microstructures.

\section{Methods}\label{sec2}

This section outlines the methodologies employed to generate a dataset from phase-field simulations, which is suitable for training a surrogate model using adjusted FNO-based neural networks. The approach is divided into 6 main parts: the formulation of the phase-field model, the numerical implementation using a semi-implicit spectral method, data generation, and storage strategy for ML purposes, and finally neural network architecture and training.

\subsection{Phase-Field Model Formulation}\label{sec2_1}

\begin{figure}[H]
    \centering
    \includegraphics[scale=.3]{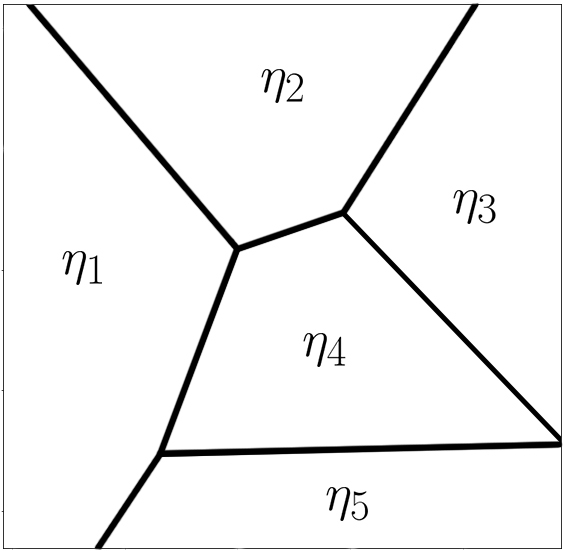}
    \caption{Illustration of order parameters in a multi-grain system using the Fan–Chen phase-field model. Each grain is represented by a distinct non-conserved order parameter $\eta_i$  that equals 1 within its respective grain and 0 elsewhere.}
    \label{fig:etas}
\end{figure}

Grain boundary movement has been widely studied using the phase-field method at the meso-scale level \cite{Miyoshi2021LargescalePS, Miyoshi2016ExtendedHM, Miyoshi2016ValidationOA, Miyoshi2018BridgingMD, Miyoshi2021LargescalePS, miyoshi2021large, Miyoshi2019LargescalePS, Tourret2017GrainGC, Takaki2019CompetitiveGD, Guo2021CompetitiveGO, Guo2020OvergrowthBA, KRILLIII20023059, PhysRevB.61.14275, fan1997computer, Krill2002ComputerSO}. In this study, we employed the widely used Fan-Chen model \cite{fan1997computer} to simulate grain evolution in 2D. The Fan–Chen model is well-established and widely used for simulating grain growth in materials science. Its widespread acceptance makes it a suitable benchmark for validating new computational frameworks. Furthermore, by choosing the Fan–Chen model for testing, we ensure that our framework is evaluated under conditions that are both representative of important physical phenomena and manageable in terms of computational complexity. In this model, the domain consists of grains with various crystallographic orientations, each represented by a distinct non-conserved order parameter \(\eta_i\), which takes the value of 1 within the \(i\)-th grain and 0 elsewhere. For example, in a five-grain system, \(\eta_2\) for the second grain would take a value of 1, while the other order parameters would be zero, i.e., \((\eta_1, \eta_2, \eta_3, \eta_4, \eta_5) = (0, 1, 0, 0, 0)\). On the grain boundaries, the order parameter smoothly transitions between 0 and 1, as shown in Fig. \ref{fig:etas} (for further details, refer to \cite{fan1997computer}).

The evolution of the order parameters, which are functions of both time and space, is governed by the AC (or Ginzburg-Landau) equation:

\[
\frac{\partial \eta_i (r, t)}{\partial t} = -L_i \frac{\delta F}{\delta \eta_i (r, t)}, \quad i = 1, 2, \ldots, p,
\]

where \(t\) is time, \(p\) is the number of grains, and \(L_i\) is the kinetic coefficient corresponding to the interface mobility. \(F\) represents the total free energy functional, which is composed of local free energy and gradient energy, defined as:

\[
F = \int \left[ f_{\text{local}}(\eta_1, \eta_2, \dots, \eta_p) + \frac{\kappa}{2} \sum_{i=1}^{p} |\nabla \eta_i|^2 \right] \, dv,
\]

where \(f_{\text{local}}\) is the local free energy and \(\kappa\) is the gradient energy coefficient. In this equation, the interfacial energy is assumed isotropic, and the local free energy density function is defined as:

\[
f(\eta_1, \eta_2, \dots, \eta_p) = \sum_{i=1}^{p} \left( -\frac{\alpha}{2} \eta_i^2 + \frac{\beta}{4} \eta_i^4 \right) + \gamma \sum_{i=1}^{p} \sum_{j \neq i}^{p} \eta_i^2 \eta_j^2,
\]

where \(\alpha\), \(\beta\), and \(\gamma\) are positive constants. By incorporating the functional derivative of the free energy into the Ginzburg-Landau equation, the evolution of the non-conserved order parameters becomes:

\[
\frac{\partial \eta_i}{\partial t} = -L \frac{\partial f}{\partial \eta_i} + L \kappa \nabla^2 \eta_i.
\]

Note that $L$ is assumed constant for all grains. This model is numerically solved to capture the evolution of order parameters and grains over time. The next section outlines the numerical method used to solve these equations.

\subsection{Numerical Implementation Using a Semi-Implicit Spectral Method}  

Spectral methods are highly effective for domains with periodic boundary conditions, a common setup in phase-field simulations. These methods achieve greater accuracy with fewer grid points, making them ideal for solving the Ginzburg-Landau equation described in section \ref{sec2_1}. To solve the equation efficiently, we apply the Fast Fourier Transform (FFT) to convert the spatial derivatives into the reciprocal space (Fourier space), as described in \cite{shen2011spectral, chen1998applications}. The FFT inherently assumes that the functions it transforms are periodic over the simulation domain. This assumption can introduce artifacts or inaccuracies if the actual boundary conditions are not periodic. To address this and align with the requirements of the FFT, we model an infinite, bulk material system by employing periodic boundary conditions. This approach ensures that no edge effects interfere with the simulation, allowing for a more realistic representation of grain growth in a bulk material.

The evolution equation is transformed into the reciprocal space using FFT, with the Fourier transform of \(\eta(r, t)\) given by:

\[
\hat{\eta}(k, t) = \int_{-\infty}^{\infty} \int_{-\infty}^{\infty} \eta(r, t) \exp(ikr) \, dx dy,
\]

where $r= (x, y)$ and \(k\) is the wave vector. The spatial derivatives in reciprocal space are computed as:

\[
\frac{\partial^n \hat{\eta}(k, t)}{\partial x^n} = (\imath k)^n \hat{\eta}(k, t), \quad \imath = \sqrt{-1}.
\]

While periodic boundary conditions are appropriate for our study, they may present challenges when modeling systems where boundaries play a significant role, as the assumption of periodicity can lead to artificial interactions between the boundaries.

The Fourier transform of both sides of the AC equation yields the following discrete form in reciprocal space:

\[
\frac{\partial \hat{\eta}_i (k, t)}{\partial t} = -L \left[ \widehat{\frac{\partial f}{\partial \eta_i (r, t)}} \right]_k - k^2 L \kappa \hat{\eta}(k, t).
\]

Here, \(k\) includes both \(k_x\) and \(k_y\) components, representing the Fourier mode. We employ a semi-implicit time-stepping scheme to enhance numerical stability and efficiency. By treating the linear term implicitly and the nonlinear term explicitly, we obtain:

\begin{equation}
    \hat{\eta}_i^{n+1} (k, t) = \frac{\hat{\eta}_i^n (k, t) - \Delta t L \left[ \widehat{\frac{\partial f}{\partial \eta_i (r, t)}} \right]_k^n}{1 + \Delta t L \kappa k^2}
    \label{eq:semi_implicit}
\end{equation}

where \(\Delta t\) is the timestep size. The FFT and inverse FFT (IFFT) functions from the Numpy library \cite{harris2020array} are used to transform the data between real and reciprocal space. This method ensures efficient computation of grain boundary evolution over time.
The parameters for solving equation \eqref{eq:semi_implicit} are set as: \(\Delta t = 5 \times 10^{-3}\), \(\kappa = 3.0\), \(L = 10\), and \(\alpha\), \(\beta\), and \(\gamma\) are all set to 1. These parameters are non-dimensional. The choice of these parameters ensures that the interfaces between grains are numerically well-resolved, allowing for accurate representation of microstructural features. Additionally, these parameters facilitate the system reaching equilibrium before the end of the simulations, ensuring that the observed grain evolution reflects a complete process. Furthermore, the boundary conditions for the simulation domain are set to be periodic in both the x- and y-directions, effectively mimicking an infinite system and eliminating edge effects.

\subsection{Data Preparation for ML Training}
Data preparation is a critical step in training ML models, particularly for complex simulations like phase-field models of grain growth. In this study, we generated a dataset from phase-field simulations of grain evolution over time. The dataset consists of 2D arrays representing the microstructure at different time steps, which are then processed to create input-output pairs suitable for the FNO model.
\subsubsection{Dataset Generation}
The phase-field simulations were conducted over a total of 8,000 time steps to capture the dynamic evolution of grain structures. However, storing the simulation results at every time step would be memory-intensive. To manage this, we saved the state of the system every 100 time steps (80 saved time steps per simulation), resulting in a dataset that sufficiently captures the microstructural evolution while being storage-efficient.

\begin{figure}[H]
    \centering
    \includegraphics[scale=.8]{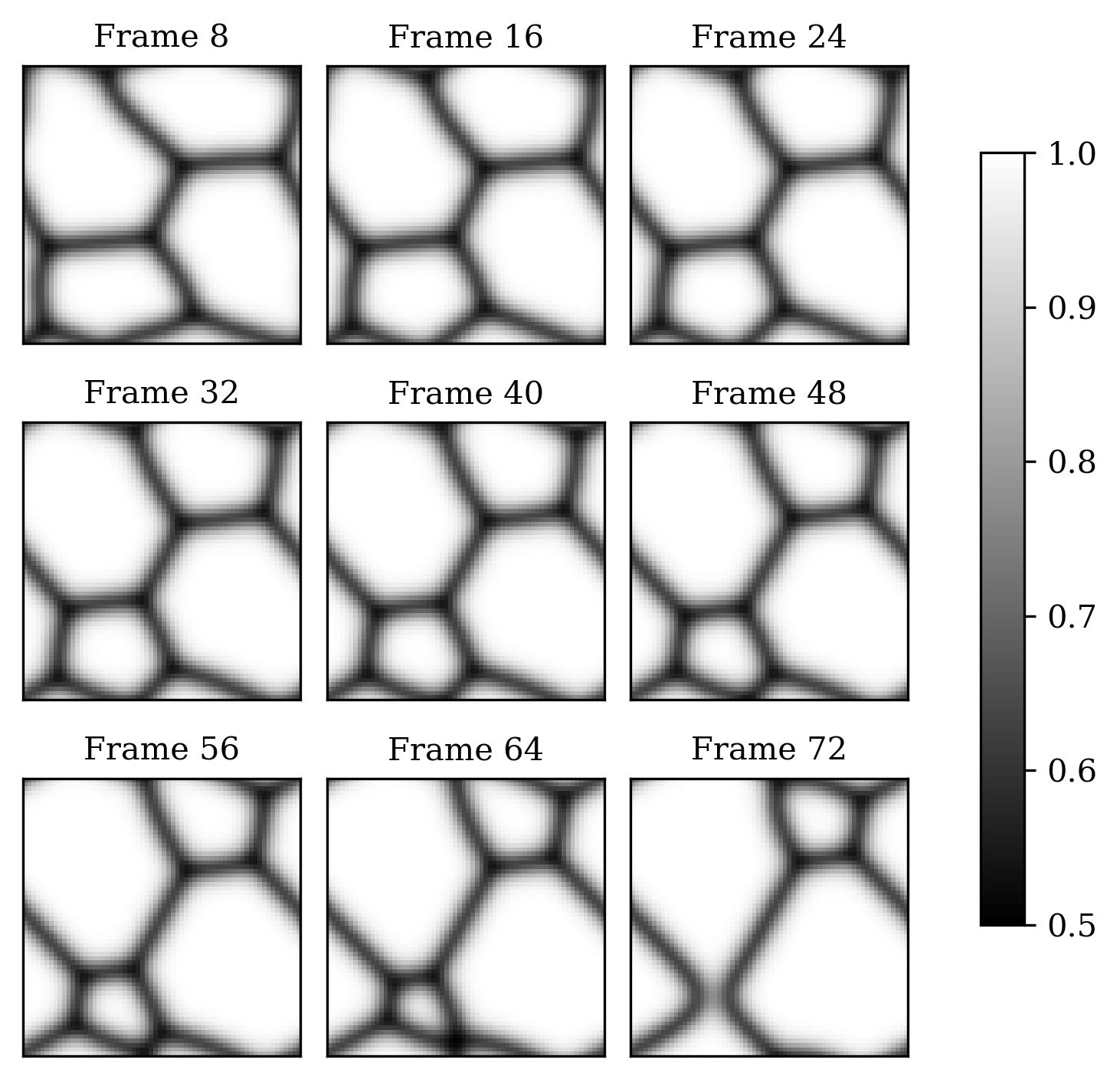}
    \caption{Example of grain evolution in a 2D phase-field simulation using the Fan–Chen model. Initial grain configurations are generated from random Voronoi diagrams and encoded using one-hot encoding. The figure displays frames at various time steps, illustrating how grains compete and evolve over time. Frame 72 shows the order parameter $\eta$ after 72,000 time steps, highlighting the microstructural changes during the simulation. Evolution of grain structure over time, illustrating the disappearance of an inner grain due to curvature-driven grain boundary migration}
    \label{fig:PF_grains}
\end{figure}

The initial values of the order parameters, \(\eta_i^0 \), define the initial grain configuration. We use the Voronoi diagram function from the SciPy package \cite{2020SciPy-NMeth} to generate random initial grain configurations. Each Voronoi plot is created using several random points, producing regions separated by straight lines. We made simulation domains with various grid point such as $64 \times 64$, $128 \times 128$ or $256 \times 256$ grid points with various spacing in both the x and y directions. After generating the Voronoi diagrams, they are converted into labeled regions. TensorFlow library \cite{tensorflow2015-whitepaper} is used to assign binary values to the order parameter arrays. In total, 1,233 simulation images were generated using random Voronoi plots, each representing a different initial microstructure and its evolution over time. 
After simulation, the arrays are transformed back into 2D arrays, where the order parameters are fixed within grains and smoothly transition across boundaries. This transformation is achieved using the argmax function from the Numpy library. Fig. \ref{fig:PF_grains} shows an example of grain evolution, where some grains compete and evolve over time. Frame 72 actually represents the order parameter ($\eta$) after 72000 time steps. While the Fan–Chen model may seem straightforward, the grain evolution it predicts is representative of fundamental mechanisms in materials science, particularly grain growth driven by the reduction of total grain boundary energy. The system tends to minimize its total free energy, specifically the grain boundary energy. Grain boundaries are regions of higher energy due to the mismatch between crystallographic orientations of adjacent grains. Reducing the total grain boundary length, lowers the system's free energy. As shown in  Fig. \ref{fig:PF_grains}, grains with higher curvature (small grains) have higher grain boundary energy and are thermodynamically less stable. They tend to shrink over time. On the other hand, grains with lower curvature (large grains) are more stable and tend to grow by consuming smaller neighboring grains. The disappearance of smaller grains is a common phenomenon during grain growth. This leads to a coarser microstructure over time. This behavior is consistent with classical theories of grain growth and has significant implications for material properties.

This prepared dataset is then used to train a neural network to predict the time evolution of microstructures based on the phase-field method, ensuring resolution-invariant learning.

\subsubsection{Data Structuring and Preprocessing}
\begin{figure}
    \centering
    \includegraphics[width=0.99\linewidth]{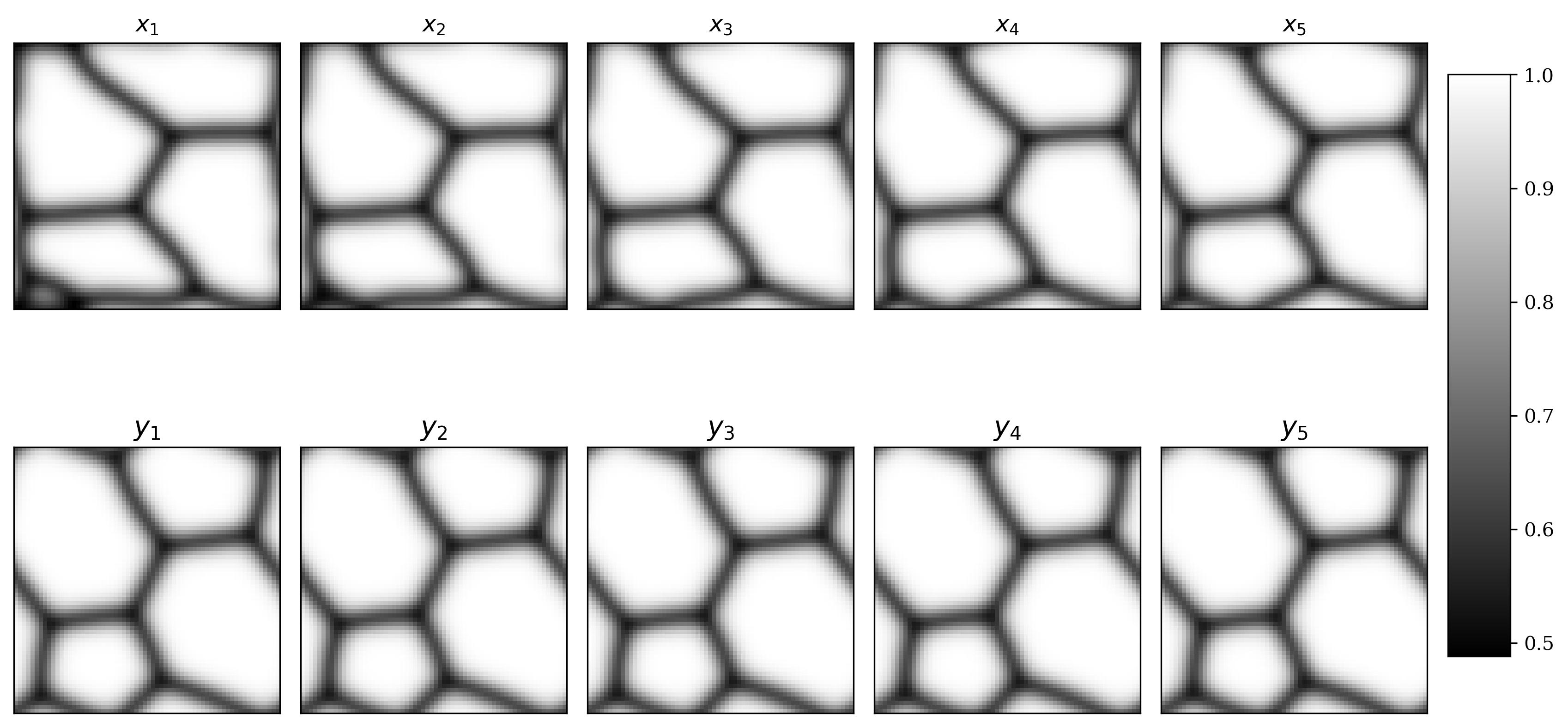}
    \caption{Examples of input (x) and output (y) microstructures used for training the neural network. The input sequence consists of microstructures from time steps \(t\) to \(t+4\), and the output sequence consists of microstructures from time steps \(t+10\) to \(t+14\), demonstrating a time shift of \(S = 10\) and a sequence length of \(T = 5\). This configuration allows the model to learn to predict future microstructures based on current and past states, facilitating the capture of temporal dependencies in microstructural evolution.}
    \label{fig:x_y}
\end{figure}

The raw dataset is a four-dimensional array with the shape \((1233, 80, 64, 64)\), where:

\begin{itemize}
    \item 1233: Number of simulation sequences.
    \item 80: Number of saved time steps per simulation (since \(8000 / 100 = 80\)).
    \item 64 $\times$ 64: Spatial dimensions of the simulation grid.
\end{itemize}
Note that we will later make datasets with different resolutions to test the framework. 
To prepare the data for training, we performed the following steps:

\begin{itemize}
    \item Time Shift for Input-Output Pairs: We aimed to predict the microstructure at future time steps based on the current and past microstructures. To achieve this, we created input-output pairs with a specified time shift. Specifically, we set a \textbf{time shift} of \(S = 10\) steps and a sequence length of \(T = 5\). For each simulation sequence: Input: Microstructures from time \(t\) to \(t + T - 1\) (i.e., times \(t, t+1, t+2, t+3, t+4\)). Output: Microstructures from time \(t + S\) to \(t + S + T - 1\) (i.e., times \(t+10, t+11, t+12, t+13, t+14\)). In other words, we used every 5 microstructures to predict the subsequent 5 microstructures after a time shift of 10 steps. The parameters \(S\) and \(T\) are adjustable, providing flexibility in the prediction horizon and the amount of historical data utilized. Example of microstructures used as input (x) and output (y) is shown in Fig. \ref{fig:x_y}

    \item Sequence Extraction with Stride: To augment the dataset and capture more temporal dynamics, we extracted overlapping sequences from each simulation run using a stride of 9 time steps. This means that for each simulation, we moved the window by 9 time steps to extract a new sequence until all possible sequences were covered.
    \item Data Cleaning: We observed that in some cases, the input and output sequences were identical, particularly when the microstructure had reached a steady state. To prevent the model from learning trivial mappings, we removed these sequences where the input equals the output within a specified tolerance level (\(1 \times 10^{-6}\)).
    \item Total Input Sequences: The preprocessed input data (\(x\)) has a shape of \((N, T_{\text{in}}, 64, 64)\), where \(N\) is the total number of sequences after data augmentation and cleaning.
  \item Total Output Sequences: The corresponding output data (\(y\)) has the same shape as the input.
\end{itemize}

Machine learning models are commonly developed and evaluated using three separate subsets of data: the training set, the validation set, and the test set. The training set is utilized to optimize the model's adjustable parameters through learning algorithms. The validation set is used to monitor the model's performance during training, helping to prevent issues like overfitting or underfitting by providing feedback on how well the model generalizes to unseen data. Finally, the test set is employed after training is complete to objectively evaluate the accuracy and predictive capability of the trained model on entirely new data \cite{peivaste2022accelerating, peivaste2023data}. We allocated 80\% of the input dataset for training the model and reserved the remaining 20\% for validation purposes. To further assess the model's generalization capabilities, we later generated separate test datasets at different resolutions. These additional test datasets allowed us to evaluate how well the trained model performs on data with resolutions it was not explicitly trained on, thereby testing its resolution-invariant properties.

\subsection{Neural Network Architecture and Training}

We make use of FNO \cite{li2020fourier} as the architecture of our framework to map solutions of the phase-field simulation. There are some advantages to using FNO, including (I) it can learn the resolution-invariant solutions, (II) this method works best in periodic boundary conditions, and this is the case for our phase-field modeling, and (III) it has shown significant accuracy in mapping solutions of highly non-linear PDEs. For more information about the theories behind FNO refer to \cite{li2020fourier, Kovachki2021NeuralOL}.
The FNO is a neural network architecture designed to learn mappings between infinite-dimensional function spaces, effectively approximating operators rather than finite-dimensional functions. This makes the FNO particularly suitable for modeling complex physical systems described by partial differential equations (PDEs). 

At the core of the FNO is the spectral convolution layer, which performs convolution operations in the Fourier space. The key components involve:

\begin{enumerate}

\item Input Lifting Layer: A fully connected layer that projects the input data into a higher-dimensional feature space. This "lifting" increases the representational capacity of the model and prepares the data for subsequent layers.

\item Fourier Transform Layers:

    \begin{itemize}
        \item FFT: The input features are transformed from the spatial domain to the frequency domain using the FFT. This operation captures global spatial patterns and long-range dependencies in the data.

        \item Convolution Layer and Spectral Convolution Layers: In the frequency domain, convolution is performed by multiplying the transformed data with learnable complex-valued weights. This process efficiently captures interactions across different scales.

        \item Inverse Fourier Transform (IFFT): After convolution in the frequency domain, the data is transformed back to the spatial domain using the inverse FFT. This ensures that the model's output remains compatible with the spatial nature of the original data.
    \end{itemize}

\item  Pointwise Nonlinear Activation Functions: After each spectral convolution layer, nonlinear activation functions are applied. These introduce nonlinearity into the model, enabling it to capture complex relationships within the data.

\item Output Projection Layers: The final output is obtained by projecting the features back to the desired dimension using fully connected layers. This maps the high-dimensional feature representations to the target output space.

\item Grid Encoding: Positional information is incorporated by concatenating normalized spatial coordinate grids to the input data. This helps the model understand spatial relationships and patterns.

\end{enumerate}

\begin{figure}
    \centering
    \includegraphics[width=0.65\linewidth]{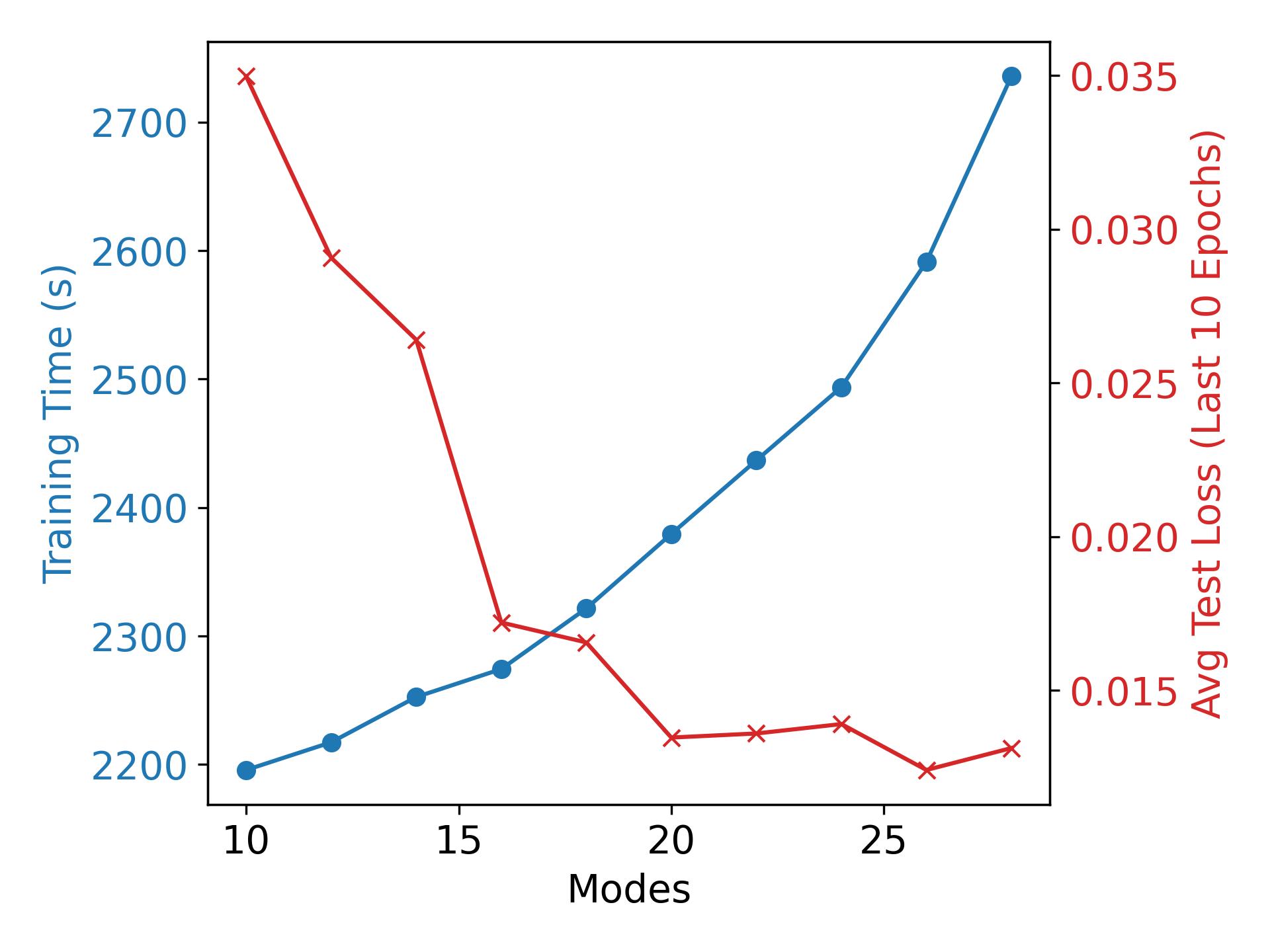}
    \caption{Impact of the number of retained Fourier modes on the model's performance and training time. The figure illustrates how varying the number of modes affects the average training loss over the last 10 epochs (blue line) and the total training time over 100 epochs (red line). As the number of modes increases, the training loss decreases due to the model's enhanced capacity to learn complex patterns, but the training time correspondingly increases because of the added computational burden. In our study, retaining 20 Fourier modes—with 2,564,917 learnable parameters—provided an optimal balance between accuracy and efficiency.}
    \label{fig:loss_mode}
\end{figure}

\begin{figure}[H]
    \centering
    \includegraphics[scale=.5]{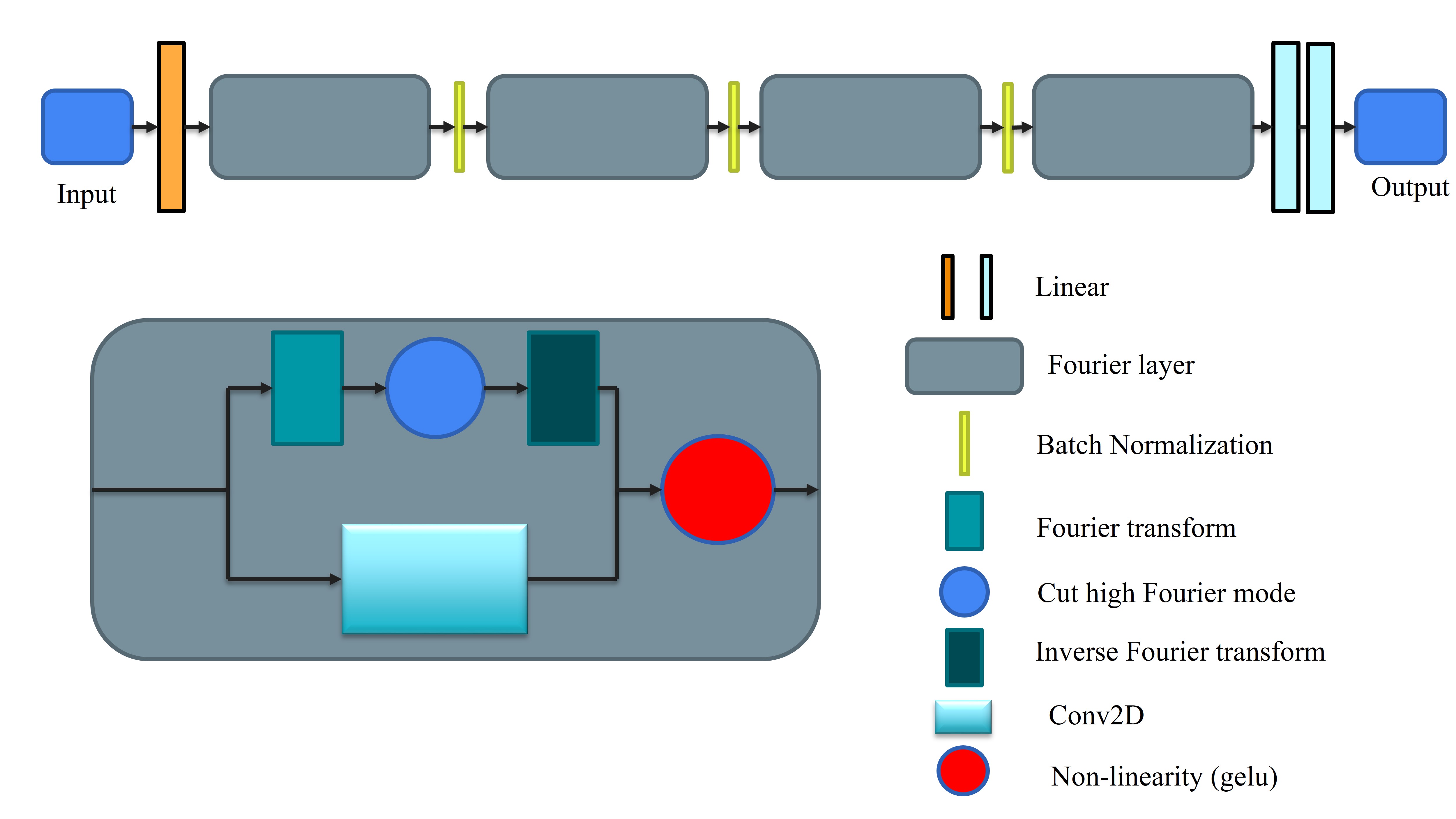}
    \caption{Architecture of the FNO used in this study. The network begins with a fully connected layer that lifts the input data to a higher-dimensional feature space with 20 channels. This is followed by four consecutive spectral convolution layers, each interleaved with pointwise convolution layers. In the spectral convolution layers, the data is transformed into the Fourier domain using the FFT. Within this domain, the network selectively retains a specified number of lower-frequency Fourier modes that capture the most significant patterns in the data, while discarding higher-frequency modes that may contribute less to the overall structure or represent noise. After mode truncation, the data is transformed back into the spatial domain using the inverse FFT. This architecture enables efficient learning of both global and local features in the microstructural evolution}
    \label{fig:FNO}
\end{figure}

The architecture of FNO used in this study is shown in Fig. \ref{fig:FNO}. Firstly, the dataset goes through a single fully connected neural network layer to increase the data dimension to 20. Four consecutive spectral convolution layers interleaved with pointwise convolution layers. 
In the spectral convolution layers of the FNO, the input data is first transformed into the Fourier domain using the FFT. In this domain, we can manipulate the data by selectively retaining a certain number of Fourier modes—specifically, the lower-frequency modes that capture the most significant patterns in the data—while discarding the higher-frequency modes that may contribute less to the overall structure or may represent noise. After this mode truncation, the data is transformed back into the real (spatial) domain using the inverse FFT.
The number of Fourier modes retained in the spectral convolution layers is a crucial hyperparameter that significantly affects the model's performance and computational efficiency. Including a higher number of modes generally allows the model to capture finer details in the data, potentially leading to lower training loss and better predictive accuracy. However, this comes at the cost of increased computational complexity, as more modes mean a larger number of learnable parameters in the model. This increase can lead to longer training times and slower inference during the testing and prediction phases.
Fig. \ref{fig:loss_mode} illustrates how varying the number of retained Fourier modes impacts both the training time over 100 epochs and the model's average loss over the last 10 epochs. The results show that as the number of modes increases, the training loss tends to decrease due to the model's enhanced capacity to learn complex patterns. However, the training time correspondingly increases because of the added computational burden. In our study, we found that using 20 Fourier modes provided an optimal balance between accuracy and efficiency. The number of learnable parameters associated with the 20 modes is 2564917. The results of Fig. \ref{fig:loss_mode} are computed by NVIDIA RTX A4000 GPU. 

In the convolution layer, the number of filters and kernels are both 20, and the stride is also (1, 1). The output of both the spectral layer and the convolution layer are concatenated. Afterward, they are passed through the non-linearity activation function, which is  GELU \cite{hendrycks2016gaussian} in our study. In the final stage, data go through two fully connected layers; the first one increases the dimension of data to 128, and the last one decreases it to 1, similar to the input.

\begin{figure}[H]
    \centering
    \includegraphics[width=0.89\linewidth]{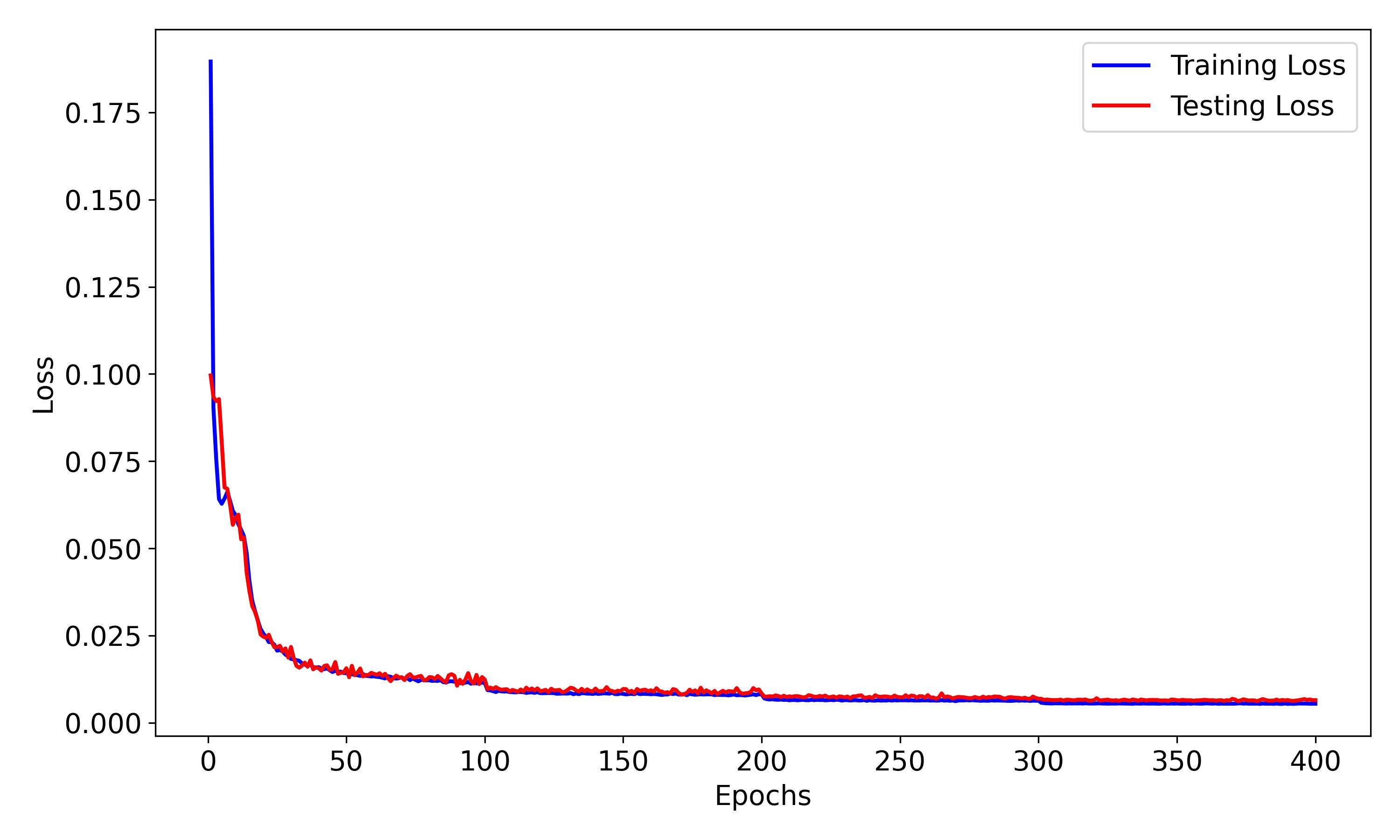}
    \caption{Training and validation loss curves over 400 epochs. The graph displays the loss values as a function of the number of epochs, showing both training and validation losses decreasing smoothly. The results indicate effective learning by the model without signs of overfitting or underfitting.}
    \label{fig:learning_curve}
\end{figure}

We divide all training data into batches of 24 sizes. The loss function used for training is the relative \(L^2\) loss, defined as:

\[
\text{Loss} = \frac{\| \hat{u} - u \|_{L^2}}{\| u \|_{L^2}},
\]

where \(\hat{u}\) is the predicted output, and \(u\) is the true output. This loss function is appropriate for regression tasks where the magnitude of the target variable can vary significantly.
The initial learning rate is 0.001; however, it gradually declines with a constant step and reduction rate during training. The ADAM optimizer \cite{kingma2014adam} is used as the gradient descent approach with $\beta_1= 0.9$, $\beta_2= 0.999$ 
and $\epsilon = 10^{-8}$. The weight decay factor is considered $10^{-4}$. 
Monitoring the loss function's value throughout network training is a common practice for evaluating the learning process. In Fig. \ref{fig:learning_curve}, we present the training and validation losses as a function of the number of epochs. The results depicted in this figure show no evidence of overfitting or underfitting. The network is trained over a total of 400 epochs. 

\section{Results and discussion}

In this section, we evaluate our model's performance on unseen microstructures from the test dataset, distinct from those used during training, to assess the network's ability to predict microstructural evolution up to 1,000 timesteps ahead. We also examine the neural network's capability to generate the entire time series by recursively using its own outputs as inputs for subsequent predictions. Additionally, we test the model on microstructures with different spatial resolutions to demonstrate the framework's effectiveness across varying scales, thereby validating its resolution-invariant properties.

\subsection{Neural Networks' Output}
\begin{figure}[H]
    \centering
    \includegraphics[width=0.99\linewidth]{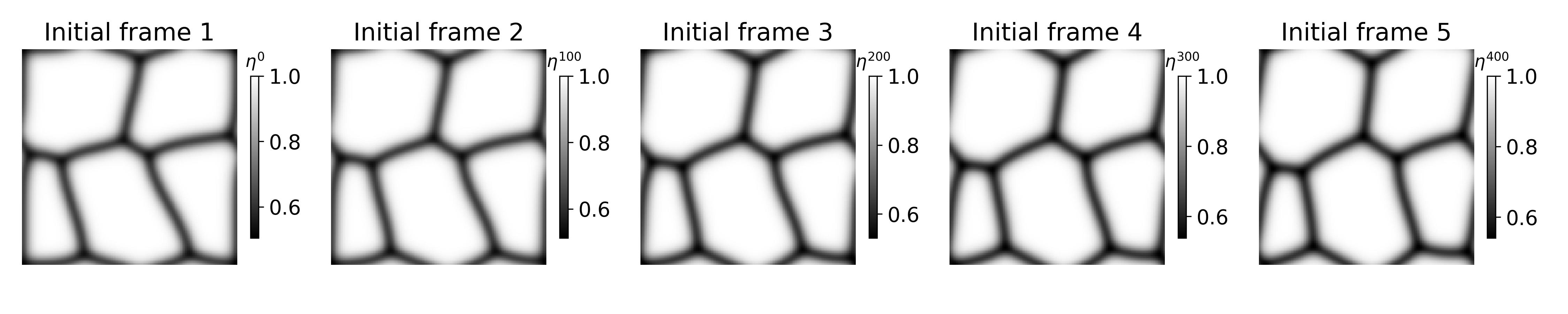}
    \caption{Initial configuration of the order parameters in an unseen microstructure used to test the neural network's generalization capability. For each prediction, the network utilizes five consecutive microstructure snapshots as input, allowing it to capture temporal dependencies and dynamic behaviors inherent in grain evolution. By incorporating historical information, the network enhances predictive accuracy and effectively models the evolution patterns of the microstructure.}
    \label{fig:initial}
\end{figure}

As outlined in our methodology, the neural network was trained to reproduce the solutions of the Fan–Chen model at any desired timestep. Fig. \ref{fig:initial} presents the initial configuration of the order parameters within a microstructure that was not included in the training dataset, the resolution of the microstructures is 128 $\times$ 128. This unseen microstructure serves as a test case to evaluate the network's generalization capability. For each prediction, the network requires five consecutive microstructure snapshots as input. By relying on multiple preceding microstructures, the network effectively captures the temporal dependencies and dynamic behaviors inherent in grain evolution. This approach enhances predictive accuracy by incorporating historical information and the evolution patterns of the microstructure.

\begin{figure}[H]
    \centering
    \includegraphics[width=0.99\linewidth]{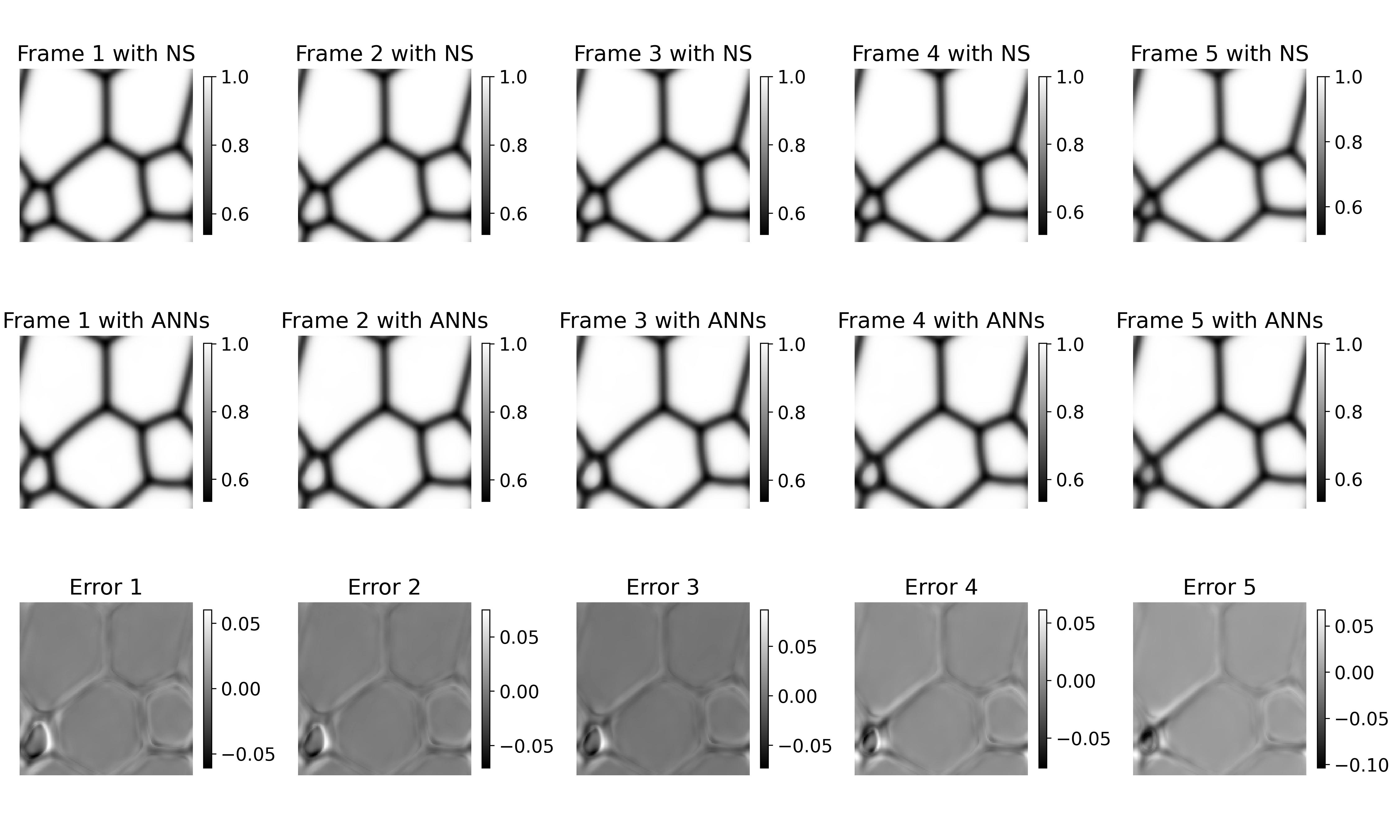}
    \caption{Evolution of the initial microstructure after 1,000 timesteps, comparing the numerical simulation (NS) and artificial neural networks (ANNs) predictions. The top row displays the results from the numerical simulation of the Fan–Chen model (ground truth), the middle row shows the corresponding outputs generated by the neural networks model, and the bottom row depicts the absolute error between the two. Despite not having seen this specific microstructure during training, the neural network effectively captures the essential features of grain evolution. The maximum error in the order parameters reaches 0.07 in localized regions, while the mean absolute error across the entire domain remains below 0.0063. The close agreement demonstrates the network's robustness and ability to generalize to unseen microstructures, accurately modeling the long-term dynamics of grain boundary evolution.}
    \label{fig:forward_pred}
\end{figure}

Fig. \ref{fig:forward_pred} illustrates the evolution of the initial configuration after 1,000 timesteps. The first row displays the results from the numerical simulation (NS) of the Fan–Chen model, which serves as the ground truth for comparison. The middle row shows the corresponding outputs generated by the artificial neural networks (ANNs). The last row depicts the absolute error between the neural network's predictions and the numerical simulation results.
The results demonstrate that neural networks' predictions closely match those obtained from the numerical solution of the Fan–Chen model. Despite the model not having seen this specific microstructure during training, it successfully captures the essential features of the grain evolution process. The error in the order parameters reaches up to 0.07 in certain localized regions of the microstructure, which may correspond to areas with complex interactions or rapid changes in grain boundaries. However, the mean absolute error across the entire domain remains below 0.0063.
This low mean absolute error indicates that the neural network, utilizing the FNO architecture, effectively models the long-term dynamics of grain boundary evolution. The ability to maintain high predictive accuracy over 1,000 timesteps showcases the network's robustness and its capacity to generalize to unseen microstructures.
The discrepancies observed in specific regions can be attributed to the inherent complexities and nonlinearities in microstructure evolution, particularly in areas with high curvature or where multiple grains interact closely. Nevertheless, the overall agreement between the neural network outputs and the numerical simulations validates the effectiveness of our approach.

\begin{figure}[H]
    \centering
    \includegraphics[width=1\linewidth]{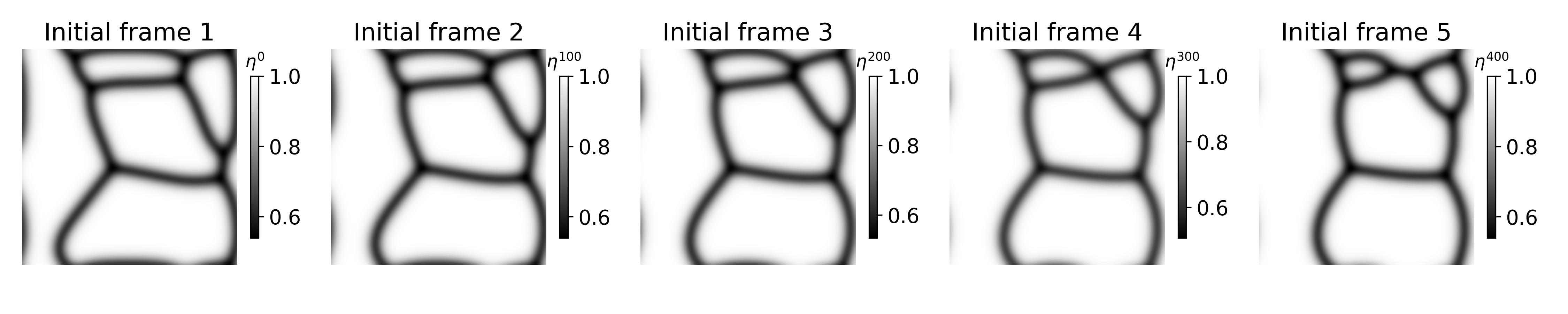}
    \caption{Initial high-resolution microstructures with a resolution of 256×256 grid points}
    \label{fig:initial_256}
\end{figure}

\begin{figure}[H]
    \centering
    \includegraphics[width=1\linewidth]{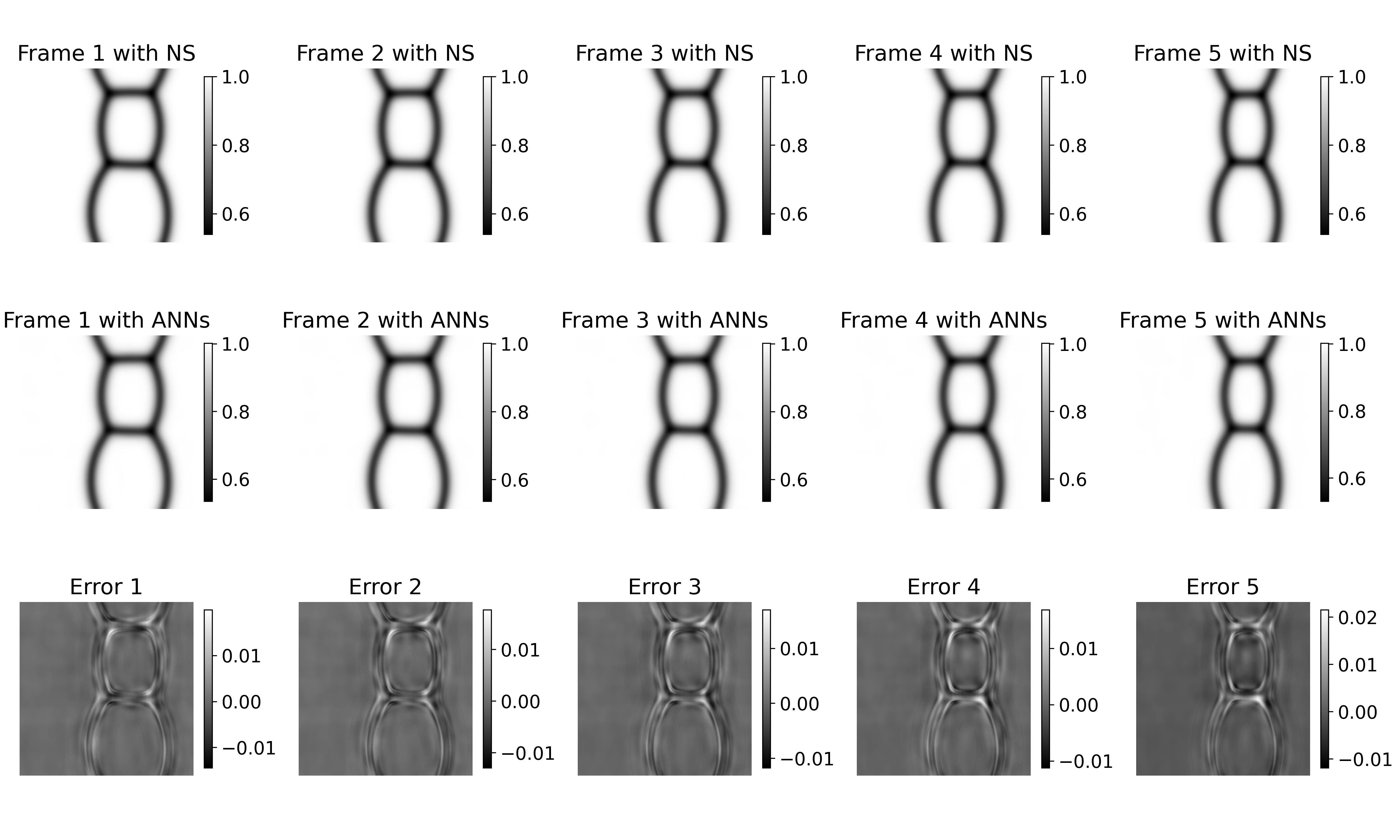}
    \caption{Comparison of grain evolution over several timesteps between NS and ANNs predictions for high-resolution microstructures (256×256 grid points). Despite the increased resolution and novel microstructural configurations not seen during training, the neural network accurately captures the dynamics of grain evolution. The maximum error between the NN predictions and NS results reaches only 0.02}
    \label{fig:256_forward}
\end{figure}

Furthermore, we evaluated the model's performance on microstructures with a higher resolution of 
256 × 256 grid points, featuring entirely new shapes that were not included in the training dataset. Fig. \ref{fig:initial_256} presents these initial high-resolution microstructures. In Fig. \ref{fig:256_forward}, we compare the grain evolution predicted by the NS and the ANNs over several timesteps.
Despite the increased resolution and the novel microstructural configurations, the neural networks accurately captured the dynamics of grain evolution. The maximum error between the neural network predictions and the numerical simulations reached only 0.02, demonstrating excellent predictive accuracy. This low error margin indicates that the model effectively generalizes to higher-resolution data and unfamiliar microstructures, showcasing its resolution-invariant capabilities.

These results further validate the effectiveness of the FNO-based approach in simulating complex microstructural phenomena across different scales. The ability to maintain high accuracy on unseen, high-resolution data underscores the model's robustness and potential for practical applications in materials science, where simulations often require handling large, detailed datasets.

\subsection{ Recursive Prediction and Long-Term Evolution}
Building on the neural network's ability to predict the microstructure at the next desired timestep with reasonable accuracy, we extended its application to recursively forecast the order parameter evolution over longer periods. By feeding the network's output back into its input, we enabled the model to simulate the microstructural dynamics over multiple timesteps without additional external input data.

\begin{figure}[H]
    \centering
    \includegraphics[width=0.45\linewidth]{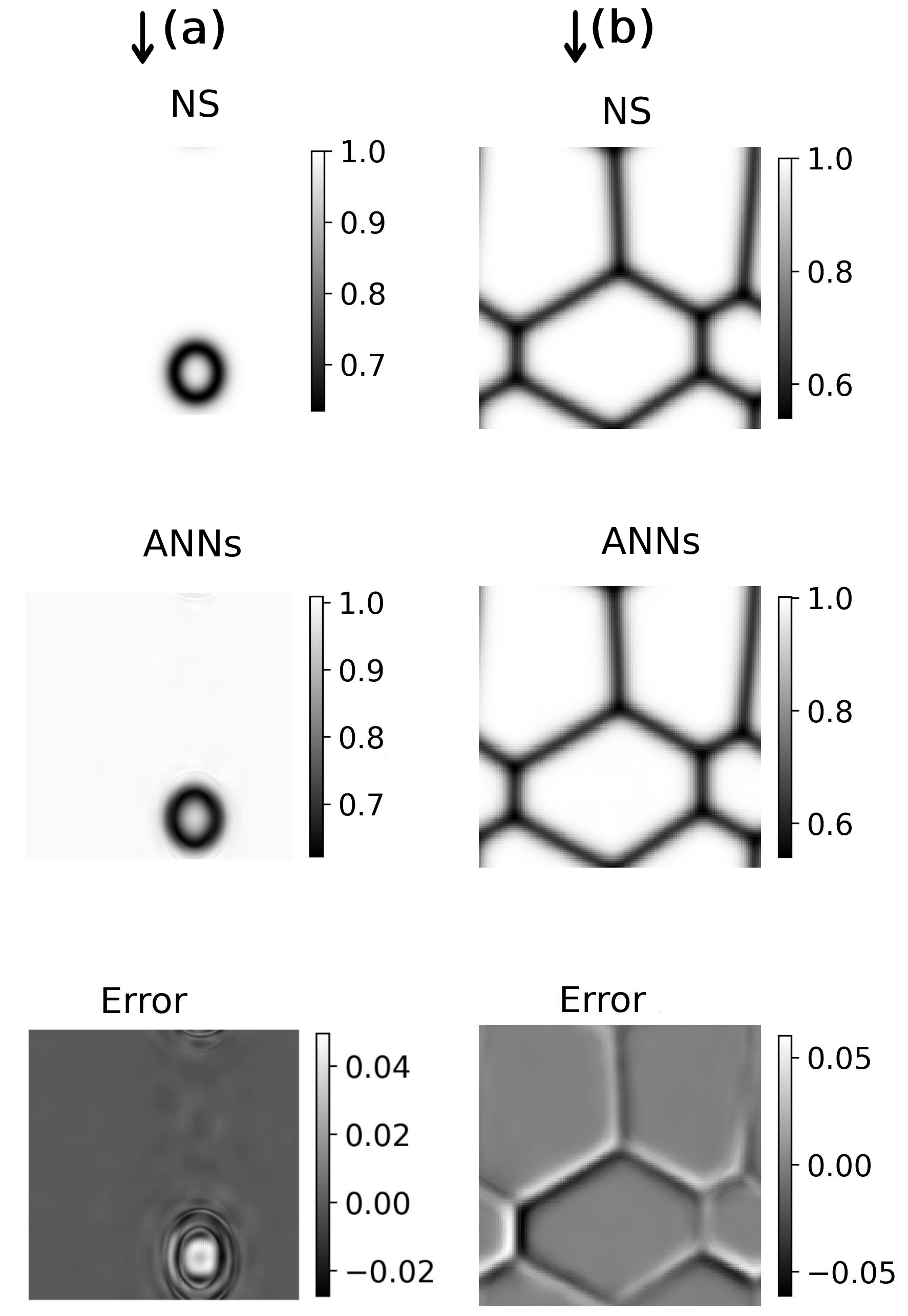}
    \caption{Temporal evolution of two distinct microstructures obtained through recursive predictions using the neural network. These microstructures were previously presented in Figs 7 and 9. In (a), the inner grain disappears over time due to grain boundary migration, (a) behavior is accurately captured by the neural networks. In (b), the grains reach a state where their areas become similar under periodic boundary conditions, resulting in minimal changes in the microstructure as time progresses. These cases demonstrate the network's capability to effectively model different grain evolution scenarios for recursive prediction}
    \label{fig:time_series}
\end{figure}

Fig. \ref{fig:time_series}a and \ref{fig:time_series}b showcase the temporal evolution of two distinct microstructures previously presented in Figs. 7 and 9. These frames represent the final outputs after recursively applying the network's predictions over several iterations. In microstructure (a), the inner grain disappears over time due to grain boundary migration, a behavior accurately captured by the neural network. In microstructure (b), the grains reach a state where their areas become similar under periodic boundary conditions, resulting in minimal changes in the microstructure as time progresses. We deliberately selected these two cases to demonstrate the network's capability to model different grain evolution scenarios effectively.
The network requires five consecutive microstructure snapshots to make each prediction. Utilizing multiple preceding states allows the model to capture the temporal dependencies and complex behaviors inherent in grain evolution processes. This approach enhances the predictive accuracy by incorporating historical information and recognizing patterns over time.
While the neural network's predictions are quite accurate, it is important to acknowledge that errors can accumulate in the recursive mode. With each successive prediction, small discrepancies may build up, potentially leading to significant deviations from the true microstructure over extended periods. This phenomenon is a common challenge in recursive prediction models, where the output at each step depends on the potentially imperfect output of the previous step.
To address this issue, one practical strategy is to periodically integrate the neural network's output with a numerical solver after a certain number of recursive predictions. Since the network outputs the order parameters directly, these can be seamlessly used as inputs for traditional numerical simulations of the Fan–Chen model. By doing so, we can correct any accumulated errors and maintain the accuracy of the microstructural evolution over longer time horizons.

\vspace{8mm}

\subsection{Computational speed}
The results presented above demonstrate that our neural network achieves high accuracy in predicting the time evolution of order parameters across a wide variety of microstructural topologies. To quantify the computational speedup obtained with the neural network relative to the conventional numerical simulation, we conducted performance evaluations using a single core of an Intel® Core™ i9-13900K CPU clocked at 3.2 GHz for both the numerical solution of the Fan–Chen model and the neural network inference.

Notably, the computation time required for evaluating microstructure evolution using the trained neural network is largely independent of the initial configuration of the order parameters. This is because the neural network inference involves a fixed set of operations determined by the network architecture, regardless of the complexity of the input microstructure. In contrast, the convergence and computational time of the numerical solver are highly dependent on the complexity of the order parameter distribution. Microstructures with intricate features or sharp gradients may require smaller time steps or more iterations to achieve convergence, thereby increasing computational time.

It is important to acknowledge that the speedup factors reported here are approximate estimates. Both the neural network inference and the FFT-based numerical calculations can be further optimized. For instance, while our focus in this study was on predicting the microstructure evolution over 500 timesteps, the neural network is capable of predicting 1,000 or even 2,000 timesteps ahead with the same computational effort per prediction due to its feedforward nature. On the other hand, the FFT-based solver could employ advanced adaptive time integration algorithms or parallel computing techniques to accelerate computations.
In our approach, the majority of the computational burden is transferred to the training phase of the neural network, which is performed offline. Once trained, the neural network can perform inference at minimal computational cost. This characteristic is particularly advantageous for applications requiring rapid predictions or real-time simulations. To further reduce the computational cost associated with data generation and model training, one can begin with simulations at lower resolutions. Due to the resolution-invariant properties of the FNO, the trained model can generalize to higher resolutions without retraining, thereby saving computational resources.

Our results indicate significant computational speedups when using the neural network for microstructure evolution predictions. Specifically, for capturing the full evolution of the microstructure depicted in Fig. 7 with a resolution of 128×128, the neural network achieved a speedup of approximately 400 times compared to the numerical solver. For the higher-resolution case shown in Fig. 9 with a resolution of 256×256, the neural network provided a speedup of around 1200 times.
These substantial speedups highlight the efficiency of the neural network approach, particularly in handling large-scale simulations where traditional numerical methods become computationally intensive. The ability of the neural network to maintain high accuracy while significantly reducing computation time demonstrates its potential for practical applications in materials science, such as real-time microstructure prediction, materials design, and high-throughput simulations.

\section{Conclusion}
In this study, we introduced a novel approach utilizing FNO to achieve resolution-invariant analysis of microstructural evolution in multi-grain systems modeled by the Fan–Chen phase-field method. The observed grain evolution in our simulations demonstrates the fundamental mechanisms of grain growth driven by the reduction of total grain boundary energy. By integrating a modified FNO architecture with phase-field simulations, we developed a surrogate model capable of accurately predicting grain boundary evolution across different spatial resolutions with significantly reduced computational costs.
Our methodology involved generating a comprehensive dataset from phase-field simulations, preparing the data for machine learning purposes, and designing an FNO-based neural network tailored to capture the complex dynamics of grain growth. The model was trained using sequences of microstructures and demonstrated remarkable accuracy in predicting long-term microstructural evolution, even for unseen configurations and higher-resolution grids not encountered during training.
Our results demonstrate not only the accuracy but also the robustness of the FNO-based model. The model consistently predicts microstructural evolution across various unseen configurations and resolutions, including higher-resolution grids and entirely new microstructural shapes not included in the training dataset. 
The results also showcased the model's ability to generalize effectively, maintaining high predictive accuracy while achieving substantial computational speedups compared to conventional numerical solvers. Specifically, the neural network can provide speedups of up to thousand times for high-resolution simulations, highlighting its potential for real-time applications and large-scale simulations in materials science.
By addressing the challenges of resolution dependence and computational inefficiency inherent in traditional phase-field modeling, our FNO-based approach offers a powerful tool for simulating microstructural phenomena. The inherent capability of FNO to handle periodic boundary conditions and capture both local and global patterns makes them particularly well-suited for materials science applications where microstructures play a critical role in determining material properties.

Future work may focus on further optimizing the neural network architecture, exploring the integration of additional physical phenomena such as stress or temperature fields, and extending the approach to three-dimensional simulations. The success of this study opens avenues for the broader application of neural operators in computational materials science, facilitating the design and optimization of materials with tailored microstructures through efficient and accurate simulations.

\clearpage
\noindent
\textbf{Data availability}: The codes regarding data generation are available in \href{https://github.com/Iman-Peivaste/PF-FNO}{github.com/Iman-Peivaste/PF-FNO}
\vspace{4mm}

\noindent
\textbf{Code availability}: 
The code used for ML in this study is open-source and accessible in \href{https://github.com/Iman-Peivaste/PF-FNO}{github.com/Iman-Peivaste/PF-FNO}.
\vspace{4mm}

\noindent
\textbf{Acknowledgement}:
This work was funded by the Luxembourg National Research Fund (FNR) through the grant PRIDE21/16758661/HYMAT.
\vspace{4mm}

\noindent\textbf{Author contributions}: I.P and S.B developed the initial concept and workflow. I.P. and A.M carried out the PF calculations and neural network training. All authors helped in the interpretation of the results. I.P. prepared the initial draft of the manuscript. S.B supervised the work. All authors discussed and contributed to preparing the final version of the manuscript.
\vspace{4mm}

\noindent\textbf{Competing interests}:
The authors declare no competing financial or non-financial interests.

\clearpage
\bibliographystyle{naturemag}. 
\bibliography{main.bbl}
\end{document}